\documentclass{article}

\usepackage{arxiv}
\usepackage{amsmath}
\usepackage{textcomp}

\usepackage[utf8]{inputenc} 
\usepackage[T1]{fontenc}    
\usepackage{hyperref}       
\usepackage{url}            
\usepackage{booktabs}       
\usepackage{amsfonts}       
\usepackage{nicefrac}       
\usepackage{microtype}      
\usepackage{graphicx}

\begin{document}

\title{Impedance-Matched Planar Metamaterial Beam Steerer for Terahertz Waves}

\author{Zinching Dang\thanks{The authors contributed equally to this work} $\,^{,}$\thanks{\texttt{dang@eit.uni-kl.de}}\,\,, Fabian Faul$^{\ast}$, Dominic Palm$^{\ast,}$\thanks{\texttt{dpalm@eit.uni-kl.de}}\,\,, Tassilo Fip, Jan Kappa, Sven Becker, and Marco Rahm \\ Department of Electrical and Computer Engineering and Research Center OPTIMAS\\ Technische Universität Kaiserslautrn\\ Kaiserslautern, Germany}

\maketitle

\begin{abstract}
Planar metamaterials with tailorable electromagnetic properties in the terahertz domain offer customized optics solutions that are needed for the development of imaging and spectroscopy systems. In particular, metamaterials carry the potential to substitute conventional bulky optics or can be basic building blocks for completely novel devices. In this respect it is advantageous that metamaterials can be devised to be impedance-matched to another material regarding their electromagnetic properties. Here, we design, fabricate and investigate a metamaterial with tailorable refractive index and impedance-matching to free space. The unit cell is comprised of two different pairs of cut-wires that provide almost independent control over the electric and magnetic response to an electromagnetic wave. For the example of a metamaterial with a uniform effective refractive index of 1.18, we experimentally demonstrate an amplitude transmission of more than $90\%$ and a reflectivity of about $5\%$ at a working frequency of $0.444\,$THz. As a more functional device, we fabricate a terahertz gradient index beam steerer with a linear refractive index gradient from 1.14 to 2.66. We numerically simulate the electromagnetic wave propagation through the beam steerer and compare the electric field amplitude of the deflected wave with the terahertz field distribution that is measured behind the beam steerer in an imaging terahertz time-domain spectroscope. The numerically simulated deflection angle of $6.1^{\circ}$ agrees well with the measured deflection angle of $5.95^{\circ}$. The measured peak amplitude transmission of the beam steerer amounts to almost $90\%$, which also agrees well with a simulated value of approximately $88\%$.
\end{abstract}


\section{Introduction} \label{sec:Intro}
In recent years, metamaterial-based gradient index (GRIN) optics \cite{Smith05} have been studied extensively, especially in the microwave regime \cite{Greegor05,Schurig06,Mei10,Meng13,Pfeiffer13,Jiang16}. Fewer works focus on the terahertz (THz) frequency domain, for which also remarkable progress has been made regarding the development of novel components, as e.g. frequency-reconfigurable metamaterial designs \cite{Chen08,Goldflam11,Ma14}, an Epsilon-Near-Zero (ENZ) metamaterial lens \cite{Pacheco-Pena17}, an all-dielectric lens \cite{Park14}, or a flexible-membrane lens \cite{Neu10}. 

The popularity of gradient index optical devices in all possible frequency ranges has not grown by chance only. In the terahertz technology for instance, conventional lenses are typically made of low-loss polymers like PTFE or TPX, which only provide a low, uniform refractive index. Since the refractive index is constant, phase differences and thus wavefront manipulation during the wave propagation in the material can only be obtained by geometric shaping of the interfaces (refraction) and spatial variation of the thickness of the material in the direction normal to the optical axis, which also impacts the geometry of the interfaces. As a result, such conventional optics is very bulky and voluminous.

GRIN devices on the other hand can implement the same functionality by influencing the phase of the propagating wave due to their spatial refractive index distribution, while being flat, thin and far less bulky. In the pursuit of creating materials with specific refractive index profiles, metamaterials have proven to provide the proper means. As a major advantage, metamaterials cannot only be devised to provide a specific spatial refractive index distribution, but also can be impedance-matched to other materials or vacuum. Such impedance-matched metamaterials are basically reflectionless in a defined frequency range and make anti-reflection coatings obsolete.

This work focuses primarily on the design, fabrication and experimental study of an impedance-matched planar metamaterial with tailorable refractive index distribution. The unit cells are composed of two different sets of cut-wire pairs, one of which responds to the electric field and the other to the magnetic field of an electromagnetic wave. By this means, the electric and magnetic properties of the metamaterial can be controlled almost independently, which is key to customize the refractive index and the impedance of the metamaterial at will. We verified the design concept by numerical and experimental means for the example of a metamaterial with uniform refractive index distribution and demonstrated an almost reflectionless terahertz gradient index beam steerer (GRIN-BS) as a more practical device. In all cases, the planar unit cell design allowed a straightforward fabrication of the metamaterials by standard photolithography.

\section{Unit Cell Design, Fabrication and Characterization}\label{sec:UnitCell}

Metamaterials allow the design of artificial electromagnetic media with customized electric and magnetic responses by careful geometric design of the unit cell structures. By this means, metamaterials can be devised to possess a tailored refractive index distribution, while a matching between the impedance of the metamaterial and free space is maintained. For proper design of such materials, it is necessary that the effective electric permittivity $\varepsilon_{\text{eff}}$ and the effective magnetic permeability $\mu_{\text{eff}}$ of the metamaterial can be changed independently, such that the effective refractive index $n$ and the relative impedance $z$ can be controlled individually according to the following equations:
\[
n = \sqrt{\varepsilon \mu}\quad (1) \quad \quad \text{and} \quad \quad \quad z = \sqrt{\frac{\mu}{\varepsilon}} \quad (2),
\]
Matching between the impedance of the metamaterial and free space is obtained, when the relative metamaterial impedance is $z = 1$. 

In the following, we present a planar metamaterial, whose unit cell consists of two cut wires with a resonant electric response and a cut-wire pair with a resonant magnetic response. As we show, these two sets of geometric structures offer the ability to design the electric and magnetic responses almost independently, which allows us to devise metamaterials with a tailored refractive index distribution and impedance-matching.

\begin{figure}
\centering
\includegraphics[width=.99\linewidth]{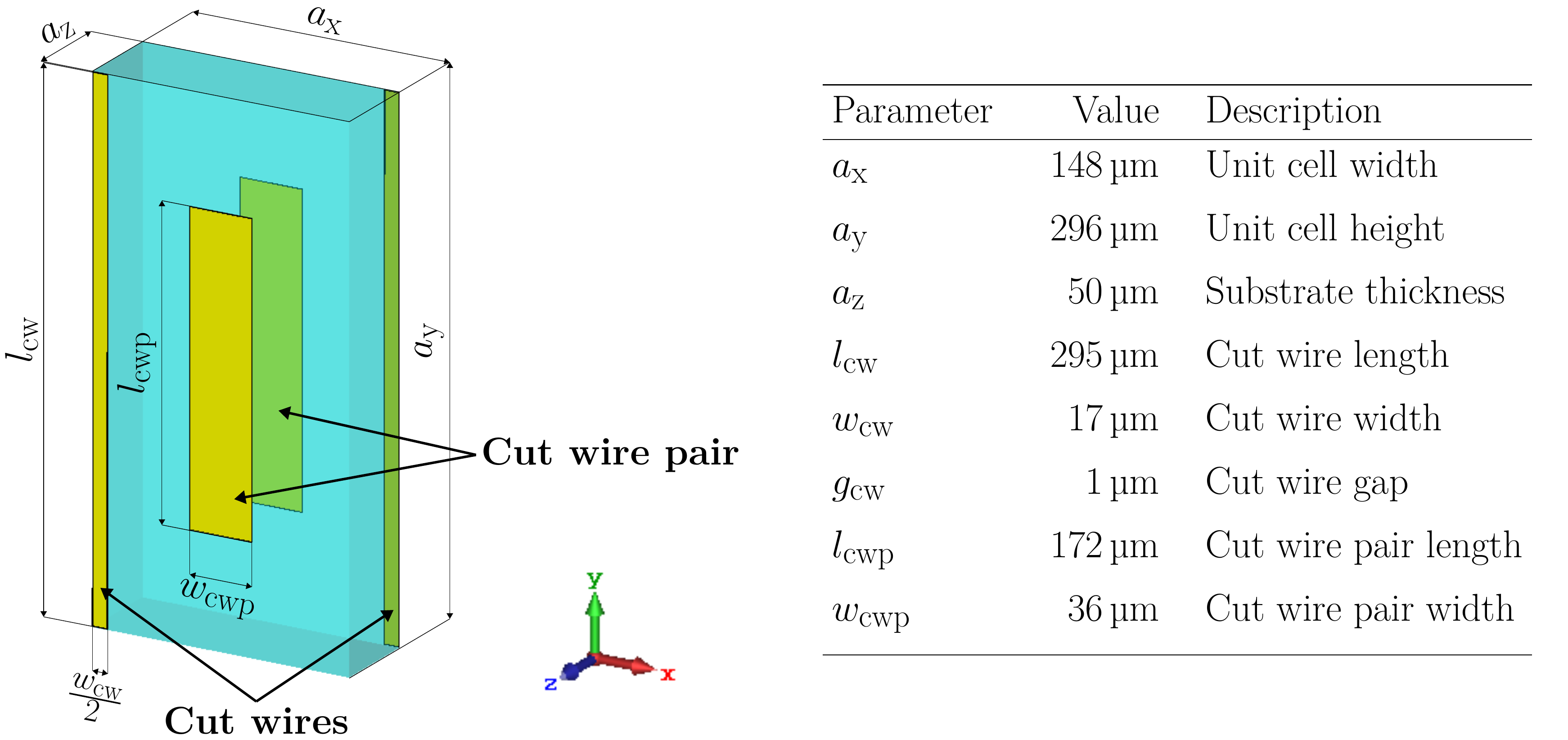}
\caption{Unit cell of an impedance-matched metamaterial with electric boundaries in $\pm y$-direction and magnetic boundaries in $\pm x$-direction. Under these conditions, the neighboring unit cells in $\pm x$-direction correspond to the mirror image of the shown unit cell at the $yz$-plane. The unit cell consists of two cut-wires at the unit cell edges (primarily electric response) and a cut-wire pair in the center of the unit cell (primarily magnetic response). Exemplary geometric parameters are given in the table right to the schematic.}\label{fig:UnitCell}
\end{figure}

Figure~\ref{fig:UnitCell} shows the schematic layout of the metamaterial unit cell from a perspective view and exemplary design values in the table. The unit cell consists of a polymer film as a substrate with metallic cut-wires on the top and bottom side. The overall dimensions are $a_x$ and $a_y$ in $x$- and $y$-direction, respectively. In $z$-direction, the thickness is $a_z + 2 \cdot t_{\text{metal}}$, where $t_{\text{metal}}$ is the thickness of the metal layer and $a_z$ is the thickness of the polymer film. Each unit cell includes four cut-wires that are arranged in two sets: two single cut-wires, each on either side at the diagonally opposing edges, and a cut-wire pair that is in the center of the unit cell. 

For the numerical simulations, we applied magnetic boundaries in $\pm x$-direction and electric boundaries in $\pm y$-direction. Under these boundary conditions and the assumption of a $y$-polarized wave, the cut-wires with a width $w_{\text{cw}}/2$ at the edges of the unit cell are mirrored at the $yz$-plane into the neighboring unit cells. As a result, the total width of the cut-wires at the unit cell edges amounts to $w_{\text{cw}}$. In the $\pm y$-direction, the unit cells are periodically continued.

As mentioned above, it is necessary to adjust the electric and magnetic response of the metamaterial independently, in order to fabricate an impedance-matched metamaterial with a deliberate refractive index distribution. In our structure, the electric response is provided by the two cut-wires at the opposite edges of the unit cell. The cut-wires couple capacitively to the electric field of a $y$-polarized electromagnetic wave. The electric resonance of the cut-wires can be tuned by changing the cut-wire width $w_{\text{cw}}$, the cut-wire length $l_{\text{cw}}$ and the cut-wire gap $g_{\text{cw}} = a_y - l_{\text{cw}}$. At the same time, the cut-wire pairs in the center of the unit cell couple inductively to the magnetic field of the $y$-polarized electromagnetic wave. In this coupling process, the magnetic resonance can be tuned independently of the electric resonance by tuning the cut-wire pair width $w_{\text{cwp}}$ and the cut-wire pair length $l_{\text{cwp}}$.

In order to verify independent control over the electric and magnetic properties in our unit cell geometry, we used CST Microwave Studio 2018\textsuperscript{\textregistered} for full-wave simulations of the electromagnetic response of the metamaterial. Figure~\ref{fig:UnitCell} shows an arbitrary example of a simulated unit cell with the corresponding geometric parameters. The metamaterial in Fig.~\ref{fig:UnitCell} was designed to offer a tailored effective refractive index $n_{\text{eff}} = 1.18$ and to provide impedance-matching to vacuum at a frequency $f_{\text{sim}} = 0.452\,$THz. As for later fabrication, the $200\,$nm thick copper wires sit on a ZeonorFilm\textsuperscript{TM} substrate with a thickness of $50\,${\textmu}m.

\begin{figure}
\centering
\includegraphics[width=.99\linewidth]{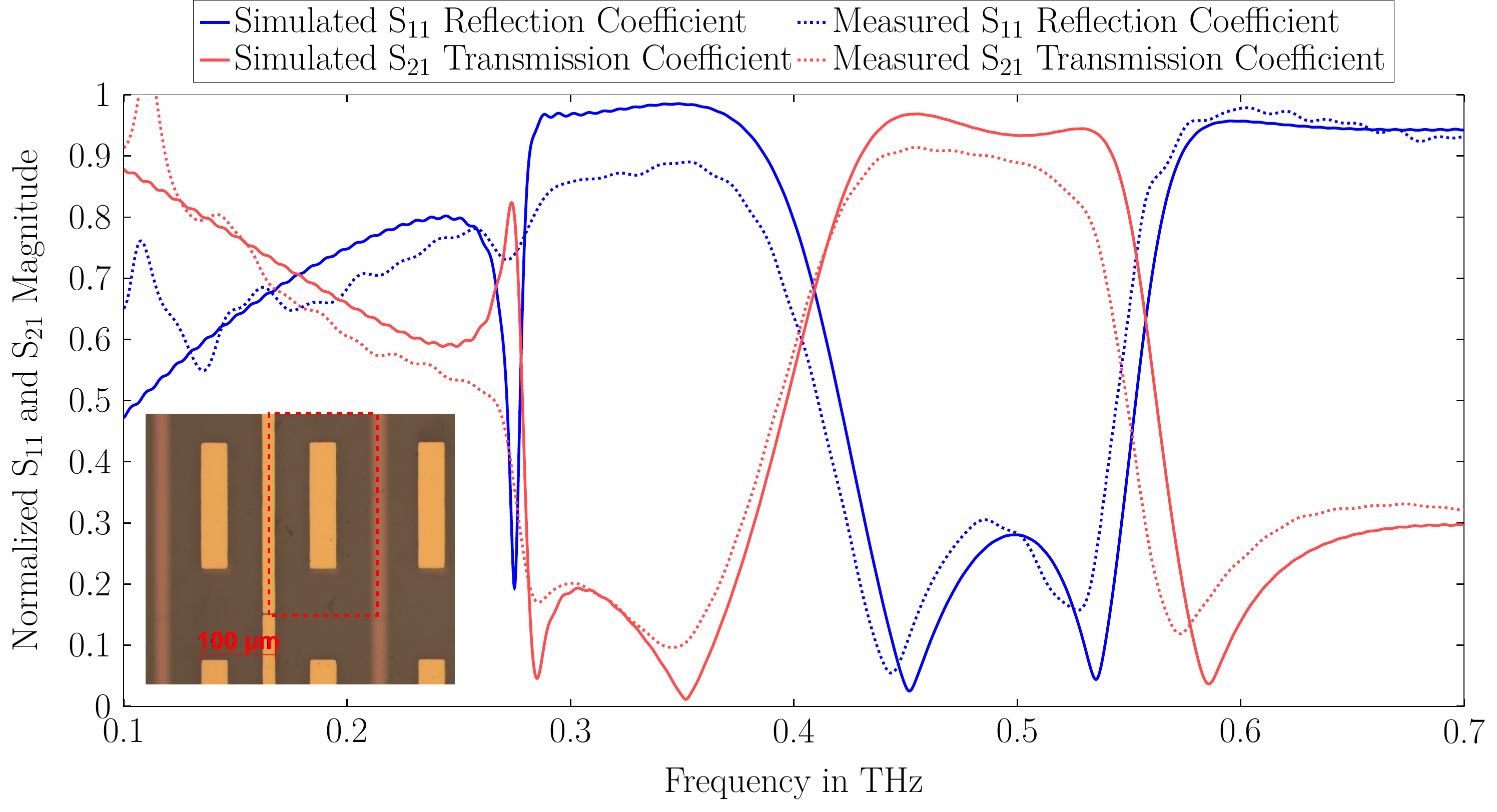}
\caption{Simulated and measured scattering parameters. The solid lines represent the simulated S$_{11}$ reflection coefficient (blue) and the S$_{21}$ transmission coefficient (red), while the dotted lines indicate the corresponding measured S$_{11}$ (blue) and S$_{21}$ (red) amplitude. The inset in the bottom left corner depicts a microscope image of the fabricated metamaterial. The red box in the inset highlights the unit cell.}\label{fig:S-Parameters}
\end{figure}

Figure~\ref{fig:S-Parameters} shows the spectral dependence of the S$_{21}$ (transmission, red solid curve) and  S$_{11}$ (reflection, blue solid curve) amplitudes for the simulated unit cell. The reflection minimum is observed at the working frequency $f_{\text{sim}} = 0.452\,$THz and is as low as $S_{11,\text{sim}} = 0.0251$, while the transmission amplitude amounts to $S_{21,\text{sim}} = 0.968$. For further quantification, we calculated the full width at half maximum (FWHM) bandwidth of the transmission band, but related to power. We determined a FWHM power bandwidth of $\text{FWHM}_{\text{sim}} = 0.146\,$THz, or $32.3\%$ with respect to the working frequency $f_{\text{sim}} = 0.452\,$THz.

\begin{figure}
\centering
\includegraphics[width=.99\linewidth, clip, trim=0 0 2cm 0]{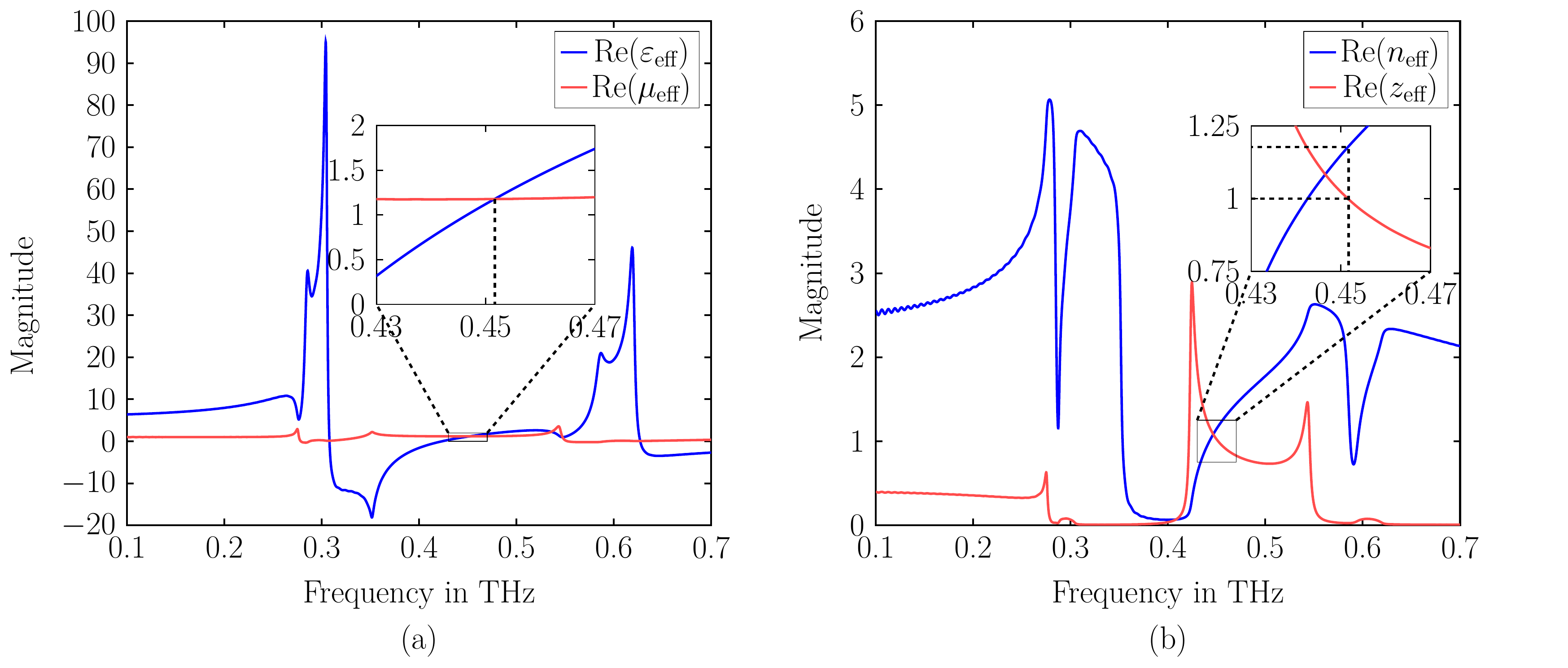}
\caption{Retrieved effective material parameters. (a) shows the real part of the permittivity $\varepsilon_{\text{eff}}$ and the permeability $\mu_{\text{eff}}$, while (b) illustrates the real part of the refractive index $n_{\text{eff}}$ and the relative wave impedance $z_{\text{eff}}$. The zoomed insets emphasize the frequency range of interest.}\label{fig:EffectiveMaterialParameters}
\end{figure}

For the calculation of the effective electromagnetic metamaterial properties, we applied the S-parameter retrieval proposed in \cite{Chen04}. Figure~\ref{fig:EffectiveMaterialParameters}(a) depicts the spectrum of the retrieved effective permittivity $\varepsilon_{\text{eff}}$ and the effective permeability $\mu_{\text{eff}}$ for the aforementioned example, whereas Figure~\ref{fig:EffectiveMaterialParameters}(b) shows the corresponding spectra of the effective refractive index $n_{\text{eff}}$ and the relative wave impedance $z_{\text{eff}}$. At the working frequency $f_{\text{sim}} = 0.452\,$THz, we observe an intersection point between the spectral curves of the permittivity and the permeability, for which both attain a value of $\varepsilon_{\text{eff}} = \mu_{\text{eff}} = 1.18$. According to Equation~(1), the metamaterial is impedance-matched to vacuum at this working frequency, which is also mirrored by the retrieved relative wave impedance of the metamaterial with a value of $z_{\text{eff}}=1$, as shown in Fig.~\ref{fig:EffectiveMaterialParameters}(b) and the zoomed image section therein. According to Equation~(2), the refractive index assumes a value $n_{\text{eff}} = 1.18$ in the impedance-matched case.

For experimental verification, we designed a metamaterial based on the preceding unit cell concept with an array of $53 \times 30$ unit cells in $x$- and $y$-direction. The aperture of the metamaterial was $7.6\,$mm$\times 8.9\,$mm. As in the simulation, we used a $50\,${\textmu}m thick ZeonorFilm\textsuperscript{TM} as a substrate, for which we determined a refractive index $n_{\text{Zeonor}} = 1.5$ by use of terahertz time-domain spectroscopy. We fabricated the cut-wires on the top and bottom surface of the substrate by UV lithography and a lift-off process. The cut-wires consist of a $190\,$nm thick copper layer on a chromium adhesion layer with $10\,$nm thickness. 

For experimental investigation of the S-parameters, we measured the amplitude of the transmission and reflection coefficients of the fabricated metamaterial in a terahertz time-domain spectroscope. The measured $S_{11}$ and $S_{21}$ spectra are plotted in Fig.~\ref{fig:S-Parameters} as dotted blue and red curve, alongside with the simulated $S_{11}$ and $S_{21}$ spectra (solid blue and solid red curve). The inset of Fig.~\ref{fig:S-Parameters} shows a microscope image section of the fabricated metamaterial. The unit cell in the inset is highlighted by the red box. The measured S-parameters agree well with the simulated S-parameters. At the working frequency $f_{\text{meas}} = 0.444\,$THz of the metamaterial, we measured a reflection coefficient minimum $S_{11,\text{meas}} = 0.054$ and a transmission coefficient $S_{21,\text{meas}} = 0.904$. In comparison with the simulated S-parameters, we observed a very slight red shift of $\Delta f_{\text{R}} = 0.008\,$THz from $f_{\text{sim}} = 0.452\,$THz to $f_{\text{meas}} = 0.444\,$THz. In addition, the measured reflection coefficient at the working frequency is by $\Delta \text{S}_{\text{11}} \approx 0.029$ higher than the corresponding simulated value, while the measured transmission coefficient is by $\Delta \text{S}_{\text{21}} \approx 0.064$ lower than the numerically calculated data. These deviations most likely stem from a combination of different effects that were not considered in the numerical calculations, as e.g. fabrication tolerances, non-perfect alignment in the measurement setup and diffraction at the finite metamaterial aperture. Nevertheless, the FWHM power bandwidth of the transmitted terahertz wave through the fabricated metamaterial is $\text{FWHM}_{\text{meas}} = 0.138\,$THz, or $31.2\%$ with respect to the working frequency $f_{\text{meas}} = 0.444\,$THz. These values agree very well with the simulated FWHM power bandwidth of $\text{FWHM}_{\text{sim}} = 0.146\,$THz, or $32.3\%$. In comparison, the difference between the experimentally and simulated FWHM power bandwidth is only $\Delta \text{FWHM} = 0.008\,$THz, or $1.1\%$. Based on these results, we conclude that our approach is well suited for the design of impedance-matched metamaterials with tailored refractive index distribution.

\section{Gradient Index Beam Steerer}\label{sec:BeamSteerer}

\begin{figure}
\centering
\includegraphics[width=.99\linewidth]{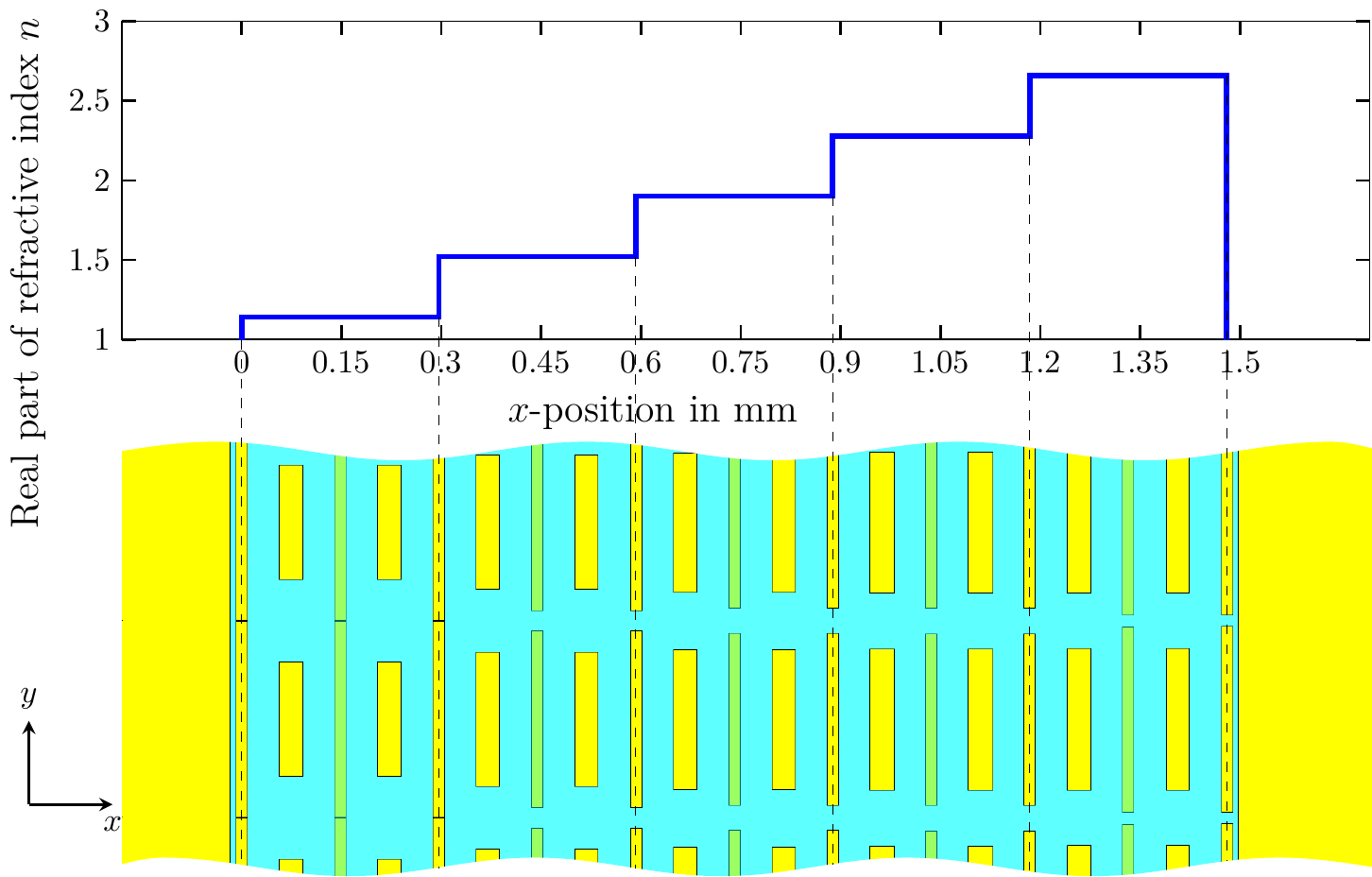}
\caption{Spatial refractive index profile and schematic of the unit cell design of the gradient index beam steerer (GRIN-BS). The GRIN-BS is framed by an aperture in $x$-direction, which is indicated by the large copper areas, and has a height of 29 unit cells in $y$-direction. Each refractive index step is implemented by 2 unit cells along the $x$-direction. The refractive index ranges from $n_1 = 1.14$ to $n_5 = 2.66$ and increases in equidistant steps of $\Delta n = 0.38$.}\label{fig:BeamSteerer}
\end{figure}

For further experimental verification of our approach, we examined a more practical example of a functional device and designed, fabricated and theoretically and experimentally investigated an impedance-matched terahertz beam steerer with linear gradient index distribution (GRIN-BS). As shown in Fig.~\ref{fig:BeamSteerer}, the spatially dependent gradient index increases linearly in five discrete steps in $x$-direction. The segment width of each step corresponds to two unit cells in $x$-direction. In $y$-direction, the metamaterial consists of 29 unit cells. By that, the aperture of the metamaterial in the $xy$-plane is divided into 5 unit cell patches with a width of $2 \times 29$ unit cells, which corresponds to an aperture of $1.48\,$mm$\times 8.58\,$mm. Each patch possesses a constant refractive index, however the refractive index increases in equidistant steps from $n_{1} = 1.14$ in the most left patch to $n_5=2.66$ in the most right patch. The spatial refractive index distribution is visualized in the upper part of Fig.~\ref{fig:BeamSteerer}. 

\begin{figure}
\centering
\includegraphics[width=.99\linewidth]{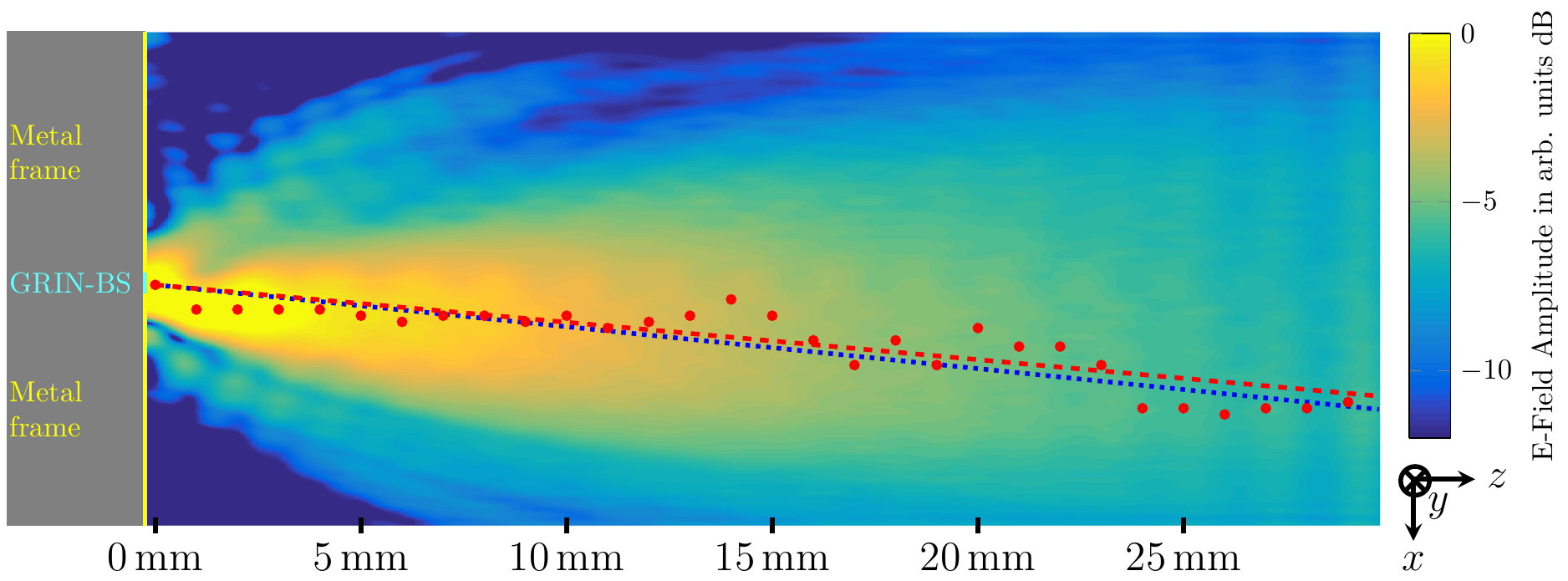}
	\caption{Interpolation of the measured spatial distribution of the terahertz electric field amplitude behind the gradient index beam steerer (GRIN-BS) at the working frequency $f_{\text{BS}} = 0.441\,$THz. The 2D map is plotted on a logarithmic scale. The white dashed line indicates the optical axis of the GRIN-BS. The blue dashed line illustrates the numerically simulated deflection angle of the GRIN-BS of $\alpha_{\text{sim}} = 6.1^{\circ}$. The red circles indicate the electric field maxima of the measured deflected terahertz wave together with the related linear fit as red dashed line. The measured deflection angle is $\alpha_{\text{meas1}} = 5.5^{\circ}$.}\label{fig:BeamSteererEField}
\end{figure}

For this linear refractive index gradient, geometric optics predicts a deflection angle of $\alpha_{\text{theo}} = 5.9^{\circ}$ for the steered beam. In order to account for diffraction, we also numerically calculated the propagation of a normally incident electromagnetic wave through the beam steerer at the working frequency $f_{\text{BS}}=0.441\,$THz. The model only accounts for diffraction that results from the finite width of the beam steerer aperture in $x$-direction, while it assumes an infinitely wide aperture in $y$-direction. The blue dashed line in Fig.~\ref{fig:BeamSteererEField} represents the trace of the electric field maxima along the propagation direction of the steered beam, from which we can deduce a deflection angle of $\alpha_{\text{sim}} = 6.1^{\circ}$. For reference, the white dashed line indicates the optical axis of the optical system. The numerically identified deflection angle is in good agreement with the theoretical calculation that neglects diffraction. The slight difference between both values can be explained by inter-cell coupling of the unit cells along the $x$-direction, which is accounted for in the numerical simulation, but not in geometric optics. In order to characterize the impedance-matching properties of the GRIN-BS, we compared the transmitted electric field amplitude $6\,$mm behind the beam steerer with the corresponding electric field amplitude after transmission through a reference aperture. We calculated a peak amplitude transmission of $88\%$ and a mean amplitude transmission of $86\%$ at the working frequency $f_{\text{BS}}=0.441\,$THz. 

For experimental demonstration of the functionality, we fabricated the designed GRIN-BS by the previously explained methods. The aperture width of the GRIN-BS was $1.48\,$mm\allowbreak $\times 8.58\,$mm. The entrance facet of the GRIN-BS was enclosed by a metallic frame to avoid leaking of incident radiation at the edges of the GRIN-BS. We measured the spatial distribution of the electric field behind the GRIN-BS by a 3-D raster scan in a terahertz time-domain spectroscope. For the amplitude transmission characterization, we measured the $xy$-plane parallel to the beam steerer, and for the deflection angle characterization, we measured the $xz$-plane along the propagation direction.

In order to characterize the amplitude transmission through the beam steerer, we measured an area of $13\,$mm$\times 10\,$mm in the $xy$-plane, $6\,$mm behind the GRIN-BS with a spatial resolution of $250\,${\textmu}m in both directions. We repeated the same procedure for the measurement of the amplitude transmission through an empty reference aperture. By comparison, we determined a peak amplitude transmission of $90\%$ and a mean amplitude transmission of $80\%$ through the beam steerer at the working frequency $f_{\text{BS}}=0.441\,$THz. The peak amplitude transmission agrees very well with the numerically calculated value, while the mean amplitude transmission is lower than predicted. The deviation in the amplitude transmission can be explained by the fabrication tolerances and imperfect polarization alignment of the incident terahertz wave. Nevertheless, the difference in the mean amplitude transmission of $6\%$ is about the same as the amplitude transmission difference of the previously characterized unit cell of $\Delta \text{S}_{\text{21}} \approx 0.064$, which corroborates our explanation.

Figure~\ref{fig:BeamSteererEField} shows the spatial distribution of the electric field amplitude behind the beam steerer in the $xz$-plane on a logarithmic scale. The electric field amplitude has been clipped at a $-12\,$dB level. The electric field amplitude was measured in an area of $30\,$mm ($z$-direction)$\times 12\,$mm ($x$-direction) behind the beam steerer. The spatial resolution of the measurement was $1\,$mm in $z$- and $150\,${\textmu}m in $x$-direction. For smoother data visualization, we interpolated the measured electric field in the illustrated $xz$ cut plane in Fig.~\ref{fig:BeamSteererEField}, however used the raw data for our analysis.

The red dots in Fig.~\ref{fig:BeamSteererEField} indicate the electric field maxima of the steered wave along the propagation direction in the $xz$-intersection plane at the working frequency $f_{\text{BS}}=0.441\,$THz. From the linear fit to the field maxima distribution (dashed red line), we determine a beam deflection angle of $\alpha_{\text{meas1}} = 5.5^{\circ}$.

Since we cannot ensure perfect normal incidence of the terahertz waves on the beam steerer, we compensated for possible misalignment by repeating the measurement with the GRIN-BS rotated by $180^{\circ}$ around the optical axis. The rotated GRIN-BS deflects the beam by a negative angle, such that the mean value of the modulus of the deflection angles for the rotated and unrotated GRIN-BS must equal the deflection angle for normal incidence. For the rotated GRIN-BS, we measured a deflection angle of $\alpha_{\text{meas2}} = 6.4^{\circ}$, which yields a mean value of $\bar{\alpha}_{\text{meas}} = 5.95^{\circ}$. In a direct comparison, the measured beam steering angle for normal incidence agrees well with the theoretical value $\alpha_{\text{theo}} = 5.9^{\circ}$ (geometric optics) and the numerically simulated value $\alpha_{\text{sim}} = 6.1^{\circ}$. From these findings, we can conclude that the refractive index distribution in the fabricated GRIN-BS matches well with the design used in the theoretical and numerical studies. A slight deviation between the measured and numerically calculated deflection angle was expected due to fabrication tolerances and the fact that the fabricated beam steerer does not elongate to infinity in $y$-direction. It was also expected that the deflection angle must be small in a one-layer design. Note that the GRIN-BS had only a thickness of $50\,${\textmu}m with a refractive index profile ranging from $1.44$ to $2.66$. By adding more layers in $z$-direction, the deflection angle can be readily increased.

\section{Conclusion}\label{sec:Conclusion}

We reported a unit cell geometry that is well suited for the design and fabrication of flexible metamaterials with tailored effective refractive index and impedance-matching to another medium, in our case vacuum. The unit cell consists of two different cut-wire pairs, one of which provides the electric response and the other the magnetic response to electromagnetic waves. The approach allows almost completely independent control of the electric and magnetic properties of the material. By means of numerical simulations, we devised impedance-matched unit cells with a refractive index between $1.14$ and $2.66$. In a first test, we fabricated a metamaterial with a uniform refractive index of 1.18 and proved an amplitude transmission of more than $90\%$ and a reflectivity as low as $5\%$ at a working frequency of $0.444\,$THz. The full width at half maximum bandwidth of the power transmission was $0.138\,$THz. In a second test, we fabricated a metamaterial-based gradient index beam steerer (GRIN-BS) and investigated its performance by numerical and experimental means. By measurement of the spatial electric field distribution of the deflected terahertz wave, we determined a deflection angle of $ 5.95^{\circ}$ in good agreement with the numerically calculated deflection angle of $6.1^{\circ}$. The measured peak amplitude transmission through the GRIN-BS was $90\%$ and the mean amplitude transmission was $80\%$. We conclude that also other terahertz components, such as lenses and wave plates, can be readily implemented by the proposed design and fabrication approach. 

\subsection*{Acknowledgment}
We acknowledge technical support by the Nano Structuring Center (NSC) at TU Kaiserslautern.

\subsection*{Funding}
This research received no external funding.

\subsection*{Conflict of interest}
The authors declare that they have no conflict of interest.

\bibliographystyle{unsrt}

\end{document}